\documentclass[showpacs,superscriptaddress,twocolumn,amsmath,aps]{revtex4} 
\usepackage{graphicx}

\newcommand{\be}{\begin{equation}}
\newcommand{\ee}{\end{equation}}
\newcommand{\bea}{\begin{eqnarray}}
\newcommand{\eea}{\end{eqnarray}}

\newcommand{\Eq}[1]{Eq.\,(\ref{#1})}

\newcommand{\la}{\langle}
\newcommand{\ra}{\rangle}
\newcommand{\dg}{\dagger}

\newcommand{\ti}{\tilde}

\newcommand{\pmthree}{\small\mbox{$\pm\frac{3}{2}$}}
\newcommand{\mthree}{\small\mbox{$-\frac{3}{2}$}}
\newcommand{\pthree}{\small\mbox{$\frac{3}{2}$}}

\newcommand{\pmone}{\small\mbox{$\pm\frac{1}{2}$}}
\newcommand{\mone}{\small\mbox{$-\frac{1}{2}$}}
\newcommand{\pone}{\small\mbox{$\frac{1}{2}$}}

\begin{document}
\draft

\title{Optical Manipulation of Single Electron Spin in Doped and
       Undoped Quantum Dots}

\author{Jinshuang Jin}
\address{State Key Laboratory for Superlattices and Microstructures,
         Institute of Semiconductors, Chinese Academy of Sciences,
         P.O. Box 912, Beijing 100083, China}
\author{Xin-Qi Li}
\address{State Key Laboratory for Superlattices and Microstructures,
         Institute of Semiconductors, Chinese Academy of Sciences,
         P.O. Box 912, Beijing 100083, China}
\affiliation{Department of Chemistry, Hong Kong University of Science and
         Technology, Kowloon, Hong Kong}
\author{YiJing Yan}
\affiliation{Department of Chemistry, Hong Kong University of Science and
         Technology, Kowloon, Hong Kong}

\date{\today}

\begin{abstract}
The optical manipulation of electron spins is of great benefit to
solid-state quantum information processing.
In this letter, we provide a comparative study on the ultrafast
optical manipulation of single electron spin in the doped and undoped
quantum dots. The study indicates that the experimental
breakthrough can be preliminarily made in the undoped quantum dots,
because of the relatively less demand.
\end{abstract}
\vspace{3ex} \pacs{ 03.67.-a, 89.70.+c, 78.47.+p, 78.67.-n}
\maketitle

In recent years quantum control of electron spins in semiconductor
nanostructures has attracted considerable attention
in the community of quantum information processing and spintronics
\cite{Wol01,Los00,Fol03,Li02,Pry02,Sha03,Los98,Ima99,Fen03,Sol03}.
In particular, the {\it all-optical} approach is greatly desirable,
since the laser pulses are much more easily controlled/tailored
than the magnetic fields, in both the time and space domains.

The basic element of ultrafast optical manipulation
of electron spins in quantum wells
has been illustrated in experiment \cite{Gup01a}.
Obviously, from the perspective of quantum device applications,
more interesting is the ultrafast manipulation of single electron spin
in quantum dots.
Several theoretical studies have focused on this issue
in {\it doped} quantum dots \cite{Li02,Pry02,Sha03}.
However, to date there are no experimental results reported in quantum dots.
In this letter, we suggest that the
illustrative experiment can be carried out preliminarily in the
{\it undoped} quantum dots, quite similar to the situation
in the {\it undoped} quantum wells \cite{Gup01a}.
This is because the controllable one-electron doped system is highly demanding.
To this purpose, a comparative study will be presented on the manipulation
in the doped and undoped quantum dots,
for both the coherent behavior and the decoherence analysis.

Let us start with the undoped quantum dot. The state diagram
is shown in Fig.\ 1(a), where
the electron in the conduction band (CB) is excited from the
valence band (VB), by using a pump laser pulse with $\sigma^{+}$-polarization
and propagating along the direction of $z$-axis.
In order to take into account the multi-electron occupation in the
VB, in this work all the electron states will be expressed in the
Fock's particle-number representation. For instance, we denote the
initially prepared state with a spin-down electron in the conduction
band by $|\psi_{0}\rangle=|0,1;1,0,1,1\rangle$, and its spin-flipped
counterpart by $|\psi_{1}\rangle=|1,0;1,0,1,1\rangle$. Here ``1"
(``0") stands for the occupation (vacancy) of the individual single
particle states, and their listing order is along, respectively, the
CB states $|\pmone\ra_c$, the VB states $|\pmthree\ra_v$ and
$|\pmone    \ra_v$.
Once the state $\psi_{0}$ as shown in Fig.\ 1(a) is initially prepared,
an off-resonance manipulating laser pulse with also the
$\sigma^+$-polarization is applied along the direction of $x$-axis.
As a result, the CB states would be virtually coupled to the VB states.
\begin{figure}\label{Fig1}
\includegraphics*[scale=0.45,angle=0.]{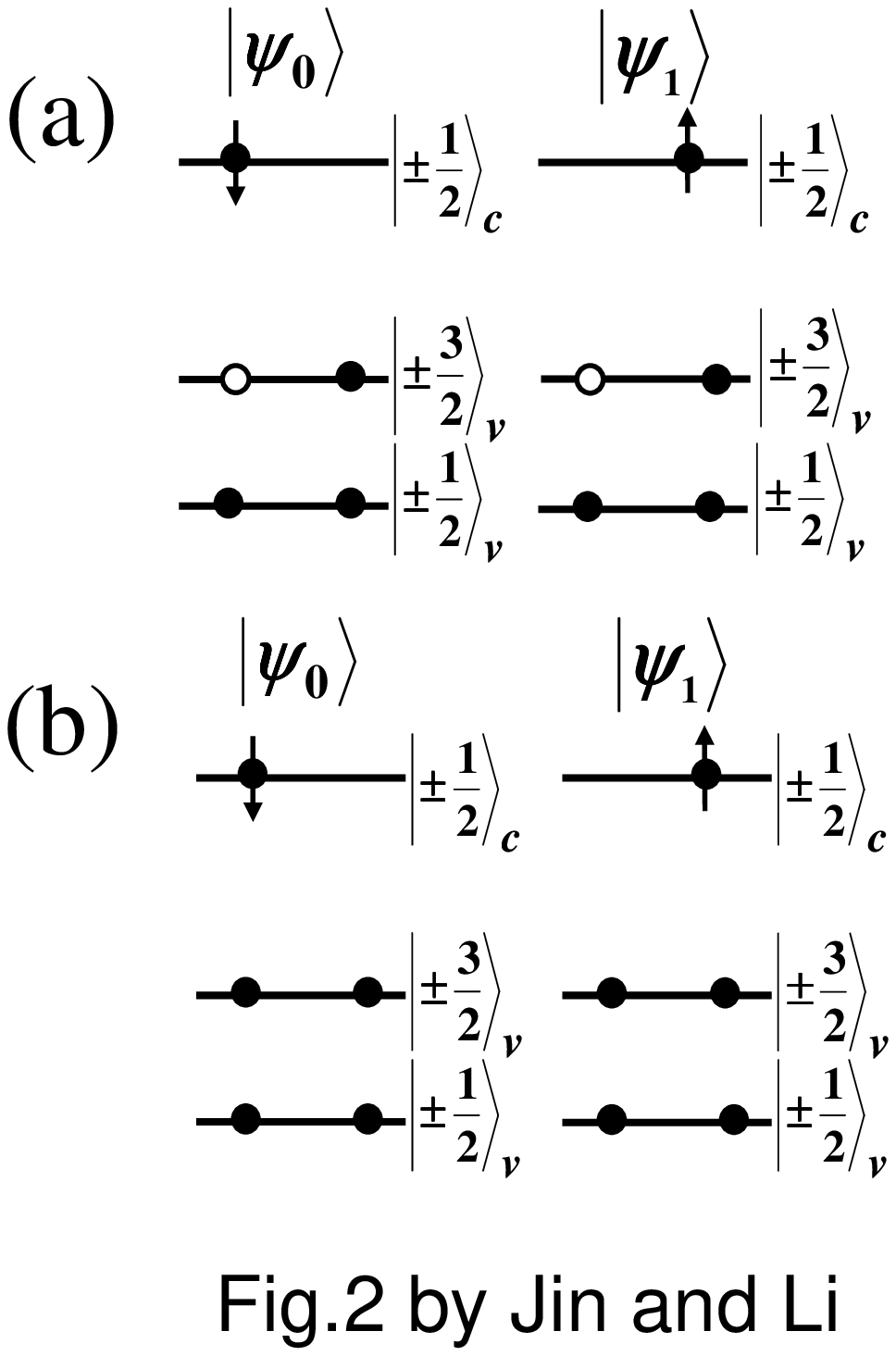}
\caption{Schematic diagram of the electron occupation, for (a) the
undoped quantum dot where an electron is excited from the valence
band to the conduction band, and (b) the doped quantum dot in which
an excess electron is injected into the conduction band.
$|\psi_0\ra$ and $|\psi_1\ra$ denote the states between which the
quantum coherent rotation is to be performed.}
\end{figure}
Moreover, an effective coupling between $|\psi_{0}\ra$ and
$|\psi_{1}\ra$ will be established by the following intermediate
virtual states in association with the transition selection rules
\cite{Bas88}:
$|\psi_{2}\rangle=|0,0;1,1,1,1\rangle$,
$|\psi_{3}\rangle=|1,1;1,0,0,1\rangle$,
$|\psi_{4}\rangle=|1,1;0,0,1,1\rangle$, and
$|\psi_{5}\rangle=|1,1;1,0,1,0\rangle$. Also, a more state
$|\psi_{\tilde{1}}\rangle=|1,0;0,1,1,1\rangle$ will be weakly
coupled to $|\psi_{0}\rangle$.
The Hilbert space spanned by these states, which we refer to as
{\it coherent subspace}, is denoted by
$\textbf{M}^{coh}=\{|\psi_{i}\rangle:i=0,1,\tilde{1},2,3,4,5\}$.
In $\textbf{M}^{coh}$, (in the interaction picture) the laser and
electron interaction Hamiltonian reads
\begin{equation}\label{Hami}
 H=\left( \begin{array}{ccccccc}
  0 & 0 & 0 & \Omega_{20}^{\ast} & \Omega_{30}^{\ast} & \Omega_{40}^{\ast}
  & \Omega_{50}^{\ast}\\
  0 & 0 & 0 & 0 & \Omega_{31}^{\ast} & 0 & \Omega_{51}^{\ast} \\
  0 & 0 & 0 & \Omega_{2\tilde{1}}^{\ast} & 0 & \Omega_{4\tilde{1}}^{\ast} & 0 \\
  \Omega_{20} & 0 & \Omega_{2\tilde{1}} & -\Delta'_{1} & 0 & 0 & 0 \\
  \Omega_{30} & \Omega_{31} & 0 & 0 & \Delta_{2} & 0 & 0 \\
  \Omega_{40}& 0 & \Omega_{4\tilde{1}} & 0 & 0 & \Delta_{1} & 0 \\
  \Omega_{50} & \Omega_{51} & 0 & 0 & 0 & 0 & \Delta_{2} \\
\end{array}
\right).
\end{equation}
\begin{figure}\label{Fig2}
\includegraphics*[scale=0.45,angle=0.]{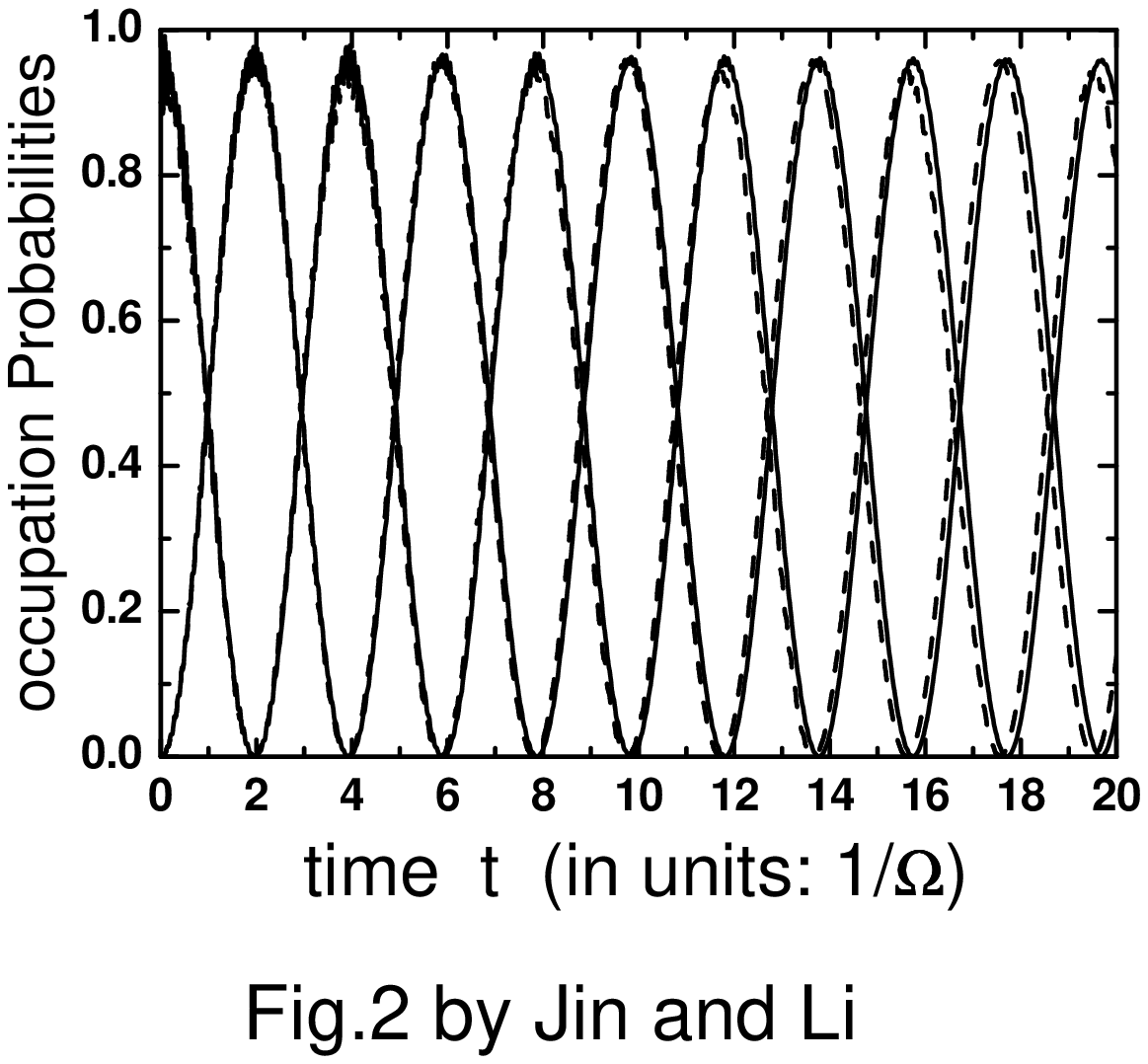}
\caption{ Laser-induced Rabi oscillation between the CB electron
spin-up and spin-down states, where the solid and dashed curves are,
respectively, for the doped and undoped quantum dots. }
\end{figure}
The matrix elements $\Omega_{ij}=eE_{0}\langle\psi_{ci}
|\vec{r}\cdot\vec{\epsilon}|\psi_{vj}\rangle$ describe the optical
coupling between the conduction and valence band states. $E_0$ and
$\vec{\epsilon}$ are, respectively, the strength and polarization of
the electric field of the laser pulse, whereas $|\psi_{ci}\rangle$
and $|\psi_{vj}\rangle$ are the single-particle CB and VB states.
Moreover, in the Hamiltonian matrix, $\Delta_{1}=E_{X1}-\hbar
\omega_p+U_{XX}$, and $\Delta_{2}=E_{X2}-\hbar \omega_p+U_{XX}$,
where $E_{X_1}$ ($E_{X_2}$) is the heavy (light) hole exciton
energy, and $U_{XX}$ is the exciton-exciton interaction energy since
two excitons appear in the intermediate states $|\psi_{3}\rangle$,
$|\psi_{4}\rangle$ and $|\psi_{5}\rangle$.
Obviously, $\Delta_{1}$ and $\Delta_{2}$ are nothing but
the detunings of the photon energy ($\hbar \omega_p$)
with respect to the excitation energies
from the heavy and light VB states to the CB state.
Finally, $\Delta'_1=\Delta_1-U_{XX}$, because $|\psi_{2}\rangle$
is the ground state which has no excitonic excitations.

By adiabatically
eliminating the intermediate states \cite{Scu97}, an effective
two-state Hamiltonian is obtained as
\begin{equation}\label{Heff}
H_{eff}= -\tilde{\Omega}(|\psi_{0}\rangle\langle\psi_{1}|
        +|\psi_{1}\rangle\langle\psi_{0}|)/2,
\end{equation}
where $\tilde{\Omega}/2=|\Omega_{31}\Omega_{30}
+\Omega_{51}\Omega_{50}|/\Delta_2$.
For an initially prepared state $|\psi_{0}\rangle$,
the above Hamiltonian will lead the state to a Rabi oscillation
between $|\psi_{0}\rangle$ and $|\psi_{1}\rangle$,
with frequency $\tilde{\Omega}/2$.
By mapping this Rabi oscillation to the classical Larmor precession
of a spin around magnetic field,
an {\it effective magnetic field} can be defined as
$\ti{B}_{eff}=\ti{\Omega}/g_e\mu_B $ \cite{Jin05,Note-1}.

Now we turn to the analysis of doped quantum dot in which an excess electron
is injected into the conduction band and the valence band is fully occupied.
The relevant energy levels are diagrammatically shown in Fig.\ 1(b).
In response to the same optical manipulation as in the undoped dot,
the electron in the VB will be virtually excited to the CB.
Using the same notation as introduced above,
the initially prepared and the spin-flipped states are, respectively,
$|\psi_{0}\rangle=|0,1;1,1,1,1\rangle$ and
$|\psi_{1}\rangle=|1,0;1,1,1,1\rangle$.
Completely similar to the analysis for the undoped quantum dot, an
effective two-state Hamiltonian in the same form of \Eq{Heff} will
be induced by a number of virtually excited intermediate states. For
the doped quantum dot, the only difference lies in the detuning
$\Delta_{2}$, which should be replaced by $\Delta_{2}'=E_{X2}-\hbar
\omega_p+U_{eX}$, with $U_{eX}$ the Coulomb interaction energy
between the doped electron and the virtually generated exciton.

Figure 2 shows the numerical result of the laser-induced Rabi oscillation
between the spin-up and spin-down states of the conduction-band electron,
for both the doped (solid curve) and undoped (dashed curve) quantum dots.
The relevant parameters are adopted as follows \cite{dot}:
$\Delta_{1}=7 \Omega$ , $\Delta_{2}=8 \Omega$,
and $U_{XX}\simeq U_{eX}=0.3\Omega$; also we set $\Omega=10$ meV.
Note that the minor difference between $U_{XX}$ and $U_{eX}$,
in particular with respect to the much larger detuning energies,
only leads to a negligibly small change of the Rabi oscillation frequency.
Strikingly, we find here an almost identical response to the coherent
optical manipulation,
for the electron spin in both the doped and undoped quantum dots.
In practice, this finding has the significant implication that the
coherent optical manipulation of single electron spin may be primarily
demonstrated in the undoped quantum dot, due to the much easier
accessibility in experiment than its doped counterpart.

In the following, we present a brief analysis for decoherence.
In semiconductor quantum dots, both the CB electron and VB hole spins
will suffer environment-induced scattering,
thus have finite decoherence times \cite{Eps01}.
In these experiments, the relatively short spin relaxation times
(with tens or hundreds of picoseconds)
may stem from the non-ideal sample preparation, the relatively large size of dots,
and/or the not low enough temperatures, etc.
From the {\it intrinsic} consideration for {\it small} quantum dots,
however, due to the large spacing between the discrete energy levels,
most spin scattering mechanisms should be strongly suppressed.
As a matter of fact, it has been
shown experimentally that in quantum dots both the electron and
hole spins are almost frozen within the electron-hole
recombination timescale, which is longer than
nanoseconds \cite{Pai01}. Also, recent theoretical calculations predicted results for
the conduction electron spin relaxation time of $10^{-6}\sim 10^{-4}$
seconds \cite{Kha01}, and hole spin relaxation time longer than
$10^{-8}$ seconds \cite{woo04}.
Therefore, in the following analysis
we only take into account the electron-hole recombination
as the {\it intrinsic} dominant decoherence source.
This treatment is reasonable at least for ultra-small quantum dots.

Based on the selection rule of optical transitions, the electron-hole
recombination is described by the jump operators
$S_{1}=|\pthree\rangle_{vc}\langle\pone|$,
$S_{2}=|\mthree\rangle_{vc} \langle\mone|$, and
$S_{3,4,5,6}=|\pmone\rangle_{vc}\langle\pmone|$.
Accordingly, the state evolution is governed by the master equation
\cite{Scu97,Lin76}
\begin{equation}\label{Lindblad}
\dot{\rho}=-i[H,\rho]-\sum^6\limits_{j=1}\gamma_{j}D[S_{j}] .
\end{equation}
The superoperator is defined by
$D[S_{j}]=\frac{1}{2}\{S^{\dg} _{j}S_{j},\rho\}-S_{j}\rho S_{j}^{\dg}$,
where $\gamma_{j}$ characterize the electron-hole recombination strengths.
Notice that under the action of the jump operators $S_j$, some
states outside the {\it coherent} subspace $\textbf{M}^{coh}$ will
be involved. Thus the state evolution described by \Eq{Lindblad} is
to be propagated in the expanded Hilbert space \cite{Jin05}.

\begin{figure}\label{Fig3}
\includegraphics*[scale=0.5,angle=0.]{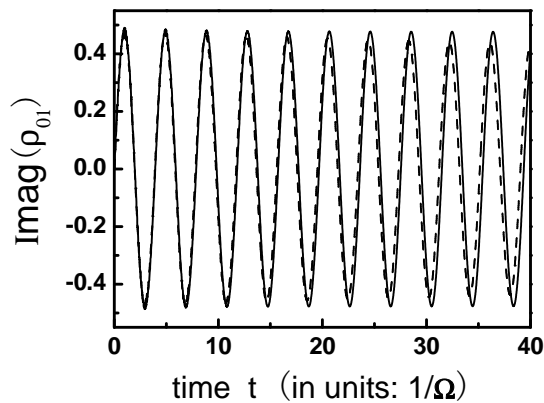}
\caption{ Effect of the electron-hole recombination on the coherent
manipulation. Plotted are, respectively, the result of the doped dot
(solid curve) and the undoped one (dashed curve). }
\end{figure}

In the numerical calculation, for the sake of simplicity we assume
identical recombination strengths, say,
$\gamma_{j}=\gamma_0=0.4\times 10^{-3}\Omega$.
The specific value adopted here is
based on the consideration that the laser-induced Rabi oscillation period
is of picoseconds and the electron-hole recombination time is of
nanoseconds \cite{Pai01}.
Other parameters are the same as used in Fig.\ 2.
Figure 3 shows the result of the doped dot (solid curve) against the
undoped one (dashed curve), in the presence of decoherence (i.e. the
electron-hole recombination). Here we plot the imaginary part of the
off-diagonal element of the density matrix, $\rho_{01}(t)=\la
\psi_0|\rho(t) |\psi_1 \ra$, which characterizes the quantum
coherence feature.
We find that in the coherent regime large number
of coherent manipulations in the undoped quantum dot can be
performed, although the manipulation in the undoped dot is less
robust against the electron-hole recombination as well as the
hole-spin relaxation than its doped counterpart \cite{Jin05}.

To summarize, we have analyzed the optical manipulation of single
electron spin in quantum dots. We found that the laser-pulse
induced effective magnetic fields are almost identical in the
doped and undoped semiconductor quantum dots. This finding
suggests that the illustrative experiment may be carried out preliminarily
in the undoped quantum dots, despite that the underlying
decoherence source such as the electron-hole recombination
will set an ultimate limit to the hundreds of
coherent spin rotations.

\vspace{3ex} {\it Acknowledgments.}
Support from the National Natural Science Foundation of China
(No.\ 90203014, 60376037, and 60425412)
and the Research Grants Council of the Hong Kong Government
is gratefully acknowledged.





\begin{references}                        
\bibitem{Wol01}
S. A. Wolf, D. D. Awschalom, R. A. Buhrman, J. M. Daughton, S. von
Molnar, M. L. Roukes, A. Y. Chtchelkanova, D. M. Treger, Science {\bf
294}, 1488 (2001).
\bibitem{Los00}                                  
Patrik Recher, Eugene V. Sukhorukov, and Daniel Loss, Phys. Rev.
Lett. {\bf 85}, 1962 (2000)
\bibitem{Fol03}                                     
J. A. Folk, R. M. Potok, C. M. Marcus, V. Umansky, Science {\bf 299},
679 (2003).
\bibitem{Li02}                                      
Xin-Qi Li, Cheng-Yong Hu, Li-Xiang Cen, and Hou-Zhi Zheng, Phys.
Rev B {\bf 66}, 235207 (2002).
\bibitem{Pry02}                                     
C. E. Pryor and M. E. Flatt$\acute{e}$, Phys. Rev. Lett. {\bf 91},
257901 (2003).
\bibitem{Sha03}                                     
Pochung Chen, C. Piermarocchi, L. J. Sham, D. Gammon, D. G. Steel,
Phys. Rev. B {\bf69}, 075320 (2004).

\bibitem{Los98}                                      
D. Loss and D. P. DiVincenzo, Phys. Rev. A {\bf 57}, 120 (1998).
\bibitem{Ima99}                                      
A. Imamoglu, D. D. Awschalom, G. Burkard, D. P. DiVincenzo, D. Loss,
M. Sherwin, and A. Small, Phys. Rev. Lett. {\bf 83}, 4204 (1999).

\bibitem{Fen03}                                     
Mang Feng, Irene D¡¯Amico, Paolo Zanardi and Fausto Rossi,
Europhysics Letts {\bf 66} 2004 14.
\bibitem{Sol03}                                      
P. Solinas, P. Zanardi, N. Zangh$\grave{i}$, and F. Rossi, Phys. Rev. A
{\bf 67}, 062315 (2003).

\bibitem{Gup01a}                            
J. A. Gupta, R. Knobel, N. Samarth, and D. D. Awschlaom, Science
{\bf 292}, 2458 (2001).

\bibitem{Bas88}
G. Bastard, {\it Wave mechanics applied to semiconductor
heterostructures} (Halsted Press, New York, 1988).
\bibitem{dot}
Here we use
$\Omega\equiv eE_0\la S|x|X\ra=eE_0\la S|y|Y\ra=eE_0\la S|z|Z\ra$
as the energy unit, where $|S\ra$, $|X\ra$, $|Y\ra$ and $|Z\ra$ are
the Bloch functions in the standard energy-band theory.

\bibitem{Jin05}
Jinshuang Jin and Xin-Qi Li, J. Appl. Phys. {\bf 98}, 123515 (2005).


\bibitem{Note-1}
We like to stress that this {\it effective magnetic field}
has no effect on the hole spin.
Under the action of the manipulating laser studied here,
only the effective coupling between the two configurations in Fig.\ 1(a)
can be established via a number of intermediate virtual states,
and the configuration with hole-spin rotation is not involved in the dynamical process.





\bibitem{Scu97}
M. O. Scully and M. S. Zubairy, {\it Quantum optics} (Cambridge
University Press, Cambridge, 1997).
\bibitem{Lin76}
G. Lindblad, Commun. Math. Phys. {\bf 48}, 199 (1976).

\bibitem{Eps01}
R. J. Epstein {et al}, Appl. Phys. Lett. {\bf78}, 733 (2001).
A. Tackeuchia {et al}, Appl. Phys. Lett. {\bf84}, 3576 (2004);
K. G$\ddot{u}$ndogdu {et al}, Appl. Phys. Lett. {\bf86}, 113111 (2005).
\bibitem{Pai01}                          
M. Paillard, X. Marie, P. Renucci, T. Amand, A. Jbeli, and J. M.
G$\acute{e}$rard, Phys. Rev. Lett. {\bf 86}, 1634 (2001).
\bibitem{Kha01}                          
A. V. Khaetskii and Y. V. Nazarov, Phys. Rev. B {\bf 61}, 12639
(2000); {\bf 64}, 125316 (2001); L. M. Woods, T. L. Reinecke, and Y.
Lyanda-Geller, Phys. Rev. B {\bf 66}, 161318 (2002); I. A.
Merkulov, Al. L. Efros, and M. Rosen, Phys. Rev. B {\bf 65}, 205309
(2002); A. V. Khaetskii, D. Loss, and L. Glazman, Phys. Rev. Lett.
{\bf 88}, 186802 (2002); J. Schliemann, A. V. Khaetskii, and D.
Loss, Phys. Rev. B {\bf 66}, 245303 (2002).
\bibitem{woo04}                          
L. M. Woods, T. L. Reinecke, and R. Kotlyar, Phys. Rev. B {\bf 69},
125330 (2004).


\end{references}
\end{document}